%% file: main.tex
\documentclass[10pt]{article}
\usepackage{amssymb,amsmath,mathrsfs,enumerate}
\usepackage{graphicx,rotate,subfigure}
\usepackage{cite}
\usepackage[colorlinks=true,
            linkcolor=red,
            urlcolor=blue,
            citecolor=blue]{hyperref}

\usepackage{color}
\long\def\rpl#1!!#2!!{\textcolor{red}{#1} \textcolor{blue}{#2}}

\def\m{\scriptstyle}
\def\l{\lambda}

\def \order(#1){{\cal O} \left(#1 \right)}

\evensidemargin=\oddsidemargin
\def\Eqn#1{Eq.\ (\ref{#1})}
\def\Eqs#1#2{Eqs.\ (\ref{#1}) and (\ref{#2})}
\allowdisplaybreaks
\begin{document}


\begin{center}
{\Large \bf Scalar sector of Two-Higgs-Doublet models: A mini-review} \\
\vspace*{1cm} {\sf Gautam Bhattacharyya\footnote{gautam.bhattacharyya@saha.ac.in}, 
~ Dipankar Das\footnote{d.das@saha.ac.in}} \\
\vspace{10pt} {\small } {\em Saha Institute of Nuclear
    Physics, 1/AF Bidhan Nagar, Kolkata 700064, India}

\normalsize
\end{center}

\begin{abstract}
A vast literature on the theory and phenomenology of Two-Higgs-Doublet
models (2HDM) exists since long. However, the present situation
demands a revisit of some 2HDM properties. Now that a 125 GeV scalar
resonance has been discovered at the LHC, with its couplings to other
particles showing increasing affinity to the Standard Model Higgs-like
behavior, the 2HDM parameter space is more squeezed than ever. We
briefly review the different parametrizations of the 2HDM potential
and discuss the constraints on the parameter space arising from the
unitarity and stability of the potential together with constraints
from the oblique electroweak $T$-parameter.  We also differentiate the
consequences of imposing a global continuous U(1) symmetry on the
potential from a discrete $Z_2$ symmetry.

\end{abstract}


\bigskip

\input{./Introduction}
\input{./Scalar_potential}

\input{./Stability}
\input{./Unitarity}
\input{./Combined_constraints}
\input{./Conclusions}


\bibliographystyle{JHEP}
\bibliography{Review.bib}
\end{document}

%% file: Introduction.tex
\section{Introduction}
The discovery of a new boson in July 2012 by the ATLAS
\cite{Aad:2012tfa} and CMS Collaborations \cite{Chatrchyan:2012ufa} of
the CERN Large Hadron Collider (LHC) is undoubtedly the greatest
achievement of this decade in the field of Particle Physics. This is
{\em most likely} `the' Higgs boson \cite{Higgs:1964pj, Higgs:1966ev,
  Englert:1964et, Guralnik:1964eu, Kibble:1967sv}, the so far eluding
final missing piece of the Standard Model (SM). But the SM has certain
inadequacies. For example, it cannot account for observations like
neutrino oscillations and dark matter. It cannot also provide adequate
matter-antimatter asymmetry of the universe. These constitute the
primary motivation to look for avenues beyond the SM, which we often
call BSM scenarios.  The SM relies on the minimal choice of a single
SU(2) scalar doublet acquiring a vacuum expectation value (vev) for
giving masses to all the particles (except the neutrino) contained in
the SM.  One natural direction towards constructing BSM scenarios is
to extend the SM scalar sector.  In doing so, one may run into the
risk of altering the tree level value of the precisely measured
oblique electroweak parameter $\rho$ (or, equivalently $T$). If we
construct an ${\rm SU(2)}\times {\rm U(1)}$ gauge theory with $N$
number of scalar multiplets, then the general expression for the tree
level $\rho$-parameter is \cite{Langacker:1980js}
\begin{eqnarray}
\rho^{\rm tree} =
\frac{\sum\limits_{i=1}^{N}\left\{T_i(T_i+1)-\frac{Y_i^2}{4}
  \right\}v_i }{\frac{1}{2} \sum\limits_{i=1}^{N}Y_i^2v_i } \,,
\end{eqnarray}
where $T_i$ and $Y_i$ denote the weak isospin and hypercharge of the
$i$-th scalar multiplet respectively, and $v_i$ is the vev acquired by
the neutral component of the that multiplet. It is easy to verify that
if the scalar sector contains only SU(2) singlets ($T_i=0$) and
doublets ($T_i=1/2$) with $Y_i = 0$ and $\pm 1$ respectively, then
$\rho^{\rm tree}=1$ is automatically satisfied without requiring any
fine tuning among the vevs. This conforms to the experimental value of
$\rho$, which is very close to unity \cite{Agashe:2014kda}. {\em In
  this article we restrict our discussions to the doublet extensions
  only}. The simplest extension of this type is two Higgs-doublet
model (2HDM) \cite{Branco:2011iw}, which has received a lot of
attention mainly because minimal supersymmetry relies on it. 2HDM
scenarios have also been investigated to look for additional sources
of CP violation for generating baryon asymmetry of the universe of
sufficient size \cite{Dorsch:2013wja}. In a general 2HDM, both the
scalar doublets, which we call $\Phi_1$ and $\Phi_2$, can couple to
fermions of both types with $T_3 = 1/2$ (up-type) and $-1/2$
(down-type).  The up- and down-type Yukawa matrices are not, in
general, simultaneously diagonalizable. This will introduce flavor
changing neutral currents (FCNC) mediated by the neutral scalars at
tree level. It was shown by Glashow and Weinberg
\cite{Glashow:1976nt}, and independently by Paschos
\cite{Paschos:1976ay}, that such tree level FCNC can be avoided if
fermions of a particular electric charge receive their masses from a
single scalar doublet. This prescription can be realized by
introducing a discrete or a continuous symmetry that apply on the
scalars $\Phi_1$ and $\Phi_2$ as well as on the fermions. Under the
discrete symmetry one of the scalars is even ($\Phi_2 \to \Phi_2$) and
the other is odd ($\Phi_1 \to - \Phi_1$).  There are four different
possibilities for assigning $Z_2$ parities to the fermions which can
avoid tree level FCNCs, i.e. ensure natural flavor conservation, so
that Glashow-Weinberg-Pashcos theorem holds true. These correspond to
the following four types of 2HDMs:
\begin{enumerate}[(i)]
\item Type I: All quarks and leptons couple to only one scalar doublet $\Phi_2$.
\item Type II: $\Phi_2$ couples to up-type quarks, while $\Phi_1$
  couples to down-type quarks and charged leptons (minimal
  supersymmetry conforms to this category).
\item Type X (or `lepton specific'): $\Phi_2$ couples to all quarks,
  while $\Phi_1$ couples to all leptons.
\item Type Y (or `flipped'): $\Phi_2$ couples to up-type quarks and
  leptons, while $\Phi_1$ couples to down-type quarks.
\end{enumerate}
There is also the option for preventing tree level FCNC by assuming
the up- and down-type Yukawa matrices to be proportional to each other
\cite{Pich:2009sp}. However, the radiative stability of the absence of
FCNC couplings in these models is not guaranteed
\cite{Ferreira:2010xe}.  The Branco-Grimus-Lavoura (BGL) model
\cite{Branco:1996bq} does on the other hand admit tree level FCNC
couplings. But those couplings are related to the off-diagonal entries
of the Cabibbo-Kobayashi-Maskawa (CKM) matrix and are naturally
suppressed. The phenomenology of the BGL scenario has been studied in
detail in Refs.~\cite{Bhattacharyya:2014nja,Botella:2014ska}.  A
conceptually similar idea for suppressing FCNC with discrete
symmetries was pursued in \cite{Joshipura:1990pi}.  In this article,
we will not elaborate any further on the FCNC issues, as we will not
discuss the Yukawa sector of 2HDM.

In the subsequent sections, we analyze the scalar potential, identify
the physical scalar eigenstates, and reach the limit in which one
physical scalar resembles the 125 GeV Higgs boson. For simplicity, we
assume all the parameters in the potential to be real so that CP is
manifestly conserved in the scalar sector.  For discussions on CP
violation in 2HDM scalar sector we refer the reader to
Refs.~\cite{Gunion:2005ja,Ferreira:2009wh,Ferreira:2010hy}, where
conditions for CP violation/conservation have been diagnosed in
detail.  With the above assumption, we derive the relations among the
parameters of the potential that need to be satisfied to ensure that
the potential is stable, i.e. it is bounded from below, at the weak
scale. We then derive the constraints arising from the requirement of
unitarity by studying the scattering amplitudes of $2 \to 2$ states
involving the scalars and the gauge bosons. After that we combine the
stability and unitarity constraints to impose numerical constraints on
the physical scalar masses and other parameters.  Now that one scalar
has been observed around 125 GeV, constraints on the remaining
parameter space have become more stringent. We conclude by
highlighting some salient features that arise from the above
considerations.

%% file: Scalar_potential.tex
\section{The scalar potential}
There are two equivalent notations that are used in the literature to
write the 2HDM scalar potential with a softly broken $Z_2$ symmetry
($\Phi_1\to \Phi_1$, $\Phi_2\to -\Phi_2$)~:
\paragraph*{\em Parametrization 1 ~:}
\begin{eqnarray}
V(\Phi_1,\Phi_2) &=& m_{11}^2 \Phi_1^\dagger\Phi_1 +
m_{22}^2\Phi_2^\dagger\Phi_2 -\left(m_{12}^2 \Phi_1^\dagger\Phi_2
+{\rm h.c.} \right) +\frac{\beta_1}{2} \left(\Phi_1^\dagger\Phi_1
\right)^2 +\frac{\beta_2}{2} \left(\Phi_2^\dagger\Phi_2 \right)^2
\nonumber \\ && +\beta_3 \left(\Phi_1^\dagger\Phi_1 \right)
\left(\Phi_2^\dagger\Phi_2 \right) +\beta_4 \left(\Phi_1^\dagger\Phi_2
\right) \left(\Phi_2^\dagger\Phi_1 \right) +\left\{\frac{\beta_5}{2}
\left(\Phi_1^\dagger\Phi_2 \right)^2 +{\rm h.c.} \right\} \,.
\label{notation1}
\end{eqnarray}
\paragraph*{\em Parametrization 2 ~:}
\begin{eqnarray}
 V &=& \lambda_1 \left( \Phi_1^\dagger\Phi_1 - \frac{v_1^2}{2}
 \right)^2 +\lambda_2 \left( \Phi_2^\dagger\Phi_2 - \frac{v_2^2}{2}
 \right)^2 +\lambda_3 \left( \Phi_1^\dagger\Phi_1 +
 \Phi_2^{\dagger}\Phi_2 - \frac{v_1^2+v_2^2}{2} \right)^2 \nonumber
 \\ && +\lambda_4 \left( (\Phi_1^{\dagger}\Phi_1)
 (\Phi_2^{\dagger}\Phi_2) - (\Phi_1^{\dagger}\Phi_2)
 (\Phi_2^{\dagger}\Phi_1) \right) + \lambda_5 \left({\rm Re}~
 \Phi_1^\dagger\Phi_2 - \frac{v_1v_2}{2} \right)^2 + \lambda_6
 \left({\rm Im}~ \Phi_1^\dagger\Phi_2 \right)^2 \,.
\label{notation2}
\end{eqnarray}
The bilinear terms proportional to $m_{12}^2$ in \Eqn{notation1} or
$\lambda_5$ in \Eqn{notation2} breaks the $Z_2$ symmetry softly.  The
presence of these soft breaking terms has implications in ensuring
decoupling behavior of these models (briefly discussed in the
concluding section).  Note that when we minimize the potential of
\Eqn{notation1}, the two minimization conditions can be used to trade
$m_{11}^2$ and $m_{22}^2$ for $v_1$ and $v_2$ and the potential can be
cast in the form of \Eqn{notation2}. The connections between the
parameters of \Eqn{notation1} and \Eqn{notation2} are given below~:
\begin{eqnarray}
&& m_{11}^2 =-(\lambda_1v_1^2+\lambda_3v^2)~;~ m_{22}^2=
  -(\lambda_2v_2^2+\lambda_3v^2)~;~
  m_{12}^2=\frac{\lambda_5}{2}v_1v_2~;~
  \beta_1=2(\lambda_1+\lambda_3)~; \nonumber \\ &&
  \beta_2=2(\lambda_2+\lambda_3)~;~ \beta_3=2\lambda_3+\lambda_4~;~
  \beta_4=\frac{\lambda_5+\lambda_6}{2}-\lambda_4~;~
  \beta_5=\frac{\lambda_5-\lambda_6}{2}~.
\label{connections}
\end{eqnarray}
In \Eqn{connections}, $v=\sqrt{v_1^2+v_2^2} = 246$ GeV, where $v_1$
and $v_2$ are the vevs of the two doublets $\Phi_1$ and $\Phi_2$
respectively. For most part of this article, we choose to work with
the notation of \Eqn{notation2}.

Before we proceed further, we comment on the relative status of the
two parametrizations of the potential. The structure in {\em
  parametrization 1} is more general than that in {\em parametrization
  2}. It is possible to go from `1' to `2' but not the other way. In
the second one it has been assumed that both scalars receive vevs,
while for the first this need not be the case. So the {\em inert
  doublet} scenario can be realized only in the first parametrization,
e.g. in a simple illustrative scenario, when the dimensionless
couplings $\beta_2 = \beta_3 = \beta_4 = \beta_5 = 0$, the mass mixing
parameter $m_{12}^2 = 0$, and $m_{22}^2 >0$. Then the second Higgs
doublet $\Phi_2$ does not acquire any vev, and the SM scalar potential
is recovered with the relation $v^2 = v_1^2 = - m_{11}^2/\beta_1$.  On
the other hand, If both doublets do indeed get vevs, there is a
correspondence between the parameters in the two cases, which we have
explicitly written down in Eq.~(\ref{connections}).

\subsection{Physical eigenstates}
We express the scalar doublets as
\begin{eqnarray}
\Phi_i =\frac{1}{\sqrt{2}} \begin{pmatrix} \sqrt{2} w_i^+ \\ (h_i+v_i)
  +iz_i \end{pmatrix} \,.
\label{phi}
\end{eqnarray}
Then we construct the mass matrices using
\Eqn{notation2}. Since we have assumed all the potential parameters to
be real, there will be no bilinear mixing term of the form
$h_iz_j$. As a result, the neutral mass eigenstates will also be the
eigenstates of CP. For the charged sector we get the following mass
matrix~:
\begin{eqnarray}
V_{\rm mass}^{\rm charged} = \begin{pmatrix} w_1^+ &
  w_2^+ \end{pmatrix} M_C^2 \begin{pmatrix} w_1^-
  \\ w_2^- \end{pmatrix} \, ,~~ {\rm with}~ M_C^2 =
\frac{\lambda_4}{2} \begin{pmatrix} v_2^2 & -v_1v_2 \\ -v_1v_2 &
  v_1^2 \end{pmatrix} \,.
\end{eqnarray}
Diagonalizing $M_C^2$ we obtain a physical charged Higgs pair
($H^\pm$) and a pair of charged Goldstones ($\omega^\pm$) as
follows~:
\begin{eqnarray}
\begin{pmatrix} \omega^\pm \\ H^\pm \end{pmatrix} = \begin{pmatrix}
  \cos\beta & \sin\beta \\ 
-\sin\beta & \cos\beta \end{pmatrix} \begin{pmatrix} w_1^\pm \\ w_2^\pm \end{pmatrix} \,,
\end{eqnarray}
where $\tan\beta=v_2/v_1$. The mass of the charged Higgs pair
($H^\pm$) is found to be
\begin{eqnarray}
m_{H^+}^2 = \frac{\lambda_4}{2}v^2 \,.
\label{m1+}
\end{eqnarray}
Similarly for the pseudoscalar part one can easily find
\begin{eqnarray}
V_{\rm mass}^{\rm CP~odd} = \begin{pmatrix} z_1 &
  z_2 \end{pmatrix}\frac{1}{2} M_P^2 \begin{pmatrix} z_1
  \\ z_2 \end{pmatrix} \, ,~~ {\rm with}~ M_P^2 =
\frac{\lambda_6}{2} \begin{pmatrix} v_2^2 & -v_1v_2 \\ -v_1v_2 &
  v_1^2 \end{pmatrix} \,.
\end{eqnarray}
The diagonalization in CP odd sector is similar to that in the charged sector. Here we
will get a physical pseudoscalar ($A$) and a neutral Goldstone
($\zeta$) as follows~:
\begin{eqnarray}
\begin{pmatrix} \zeta \\ A \end{pmatrix} = \begin{pmatrix} \cos\beta &
  \sin\beta \\ 
-\sin\beta & \cos\beta \end{pmatrix} \begin{pmatrix} z_1 \\ z_2 \end{pmatrix} \,.
\end{eqnarray}
The mass of the pseudoscalar is given by
\begin{eqnarray}
m_{A}^2 = \frac{\lambda_6}{2}v^2 \,.
\label{mA}
\end{eqnarray}
For the CP-even scalar part we find
\begin{subequations}
\begin{eqnarray}
&& V_{\rm mass}^{\rm CP~even} = \begin{pmatrix} h_1 &
    h_2 \end{pmatrix}\frac{1}{2} M_S^2 \begin{pmatrix} h_1
    \\ h_2 \end{pmatrix}~~ {\rm with,}~ M_S^2 = \begin{pmatrix} A_S &
    B_S \\ B_S & C_S \end{pmatrix} \,, \\ {\rm where,} && A_S =
  2(\lambda_1+\lambda_3)v_1^2 +\frac{\lambda_5}{2}v_2^2 \,, \\ && B_S
  =2(\lambda_3+\frac{\lambda_5}{4})v_1v_2 \,, \\ && C_S =
  2(\lambda_2+\lambda_3)v_2^2 +\frac{\lambda_5}{2}v_1^2 \,.
\end{eqnarray}
\end{subequations}
The masses of the physical eigenstates, $H$ (heavier) and $h$
(lighter), can be readily obtained as
\begin{subequations}
\begin{eqnarray}
m_{H}^2 &=& \frac{1}{2}\left[(A_S+C_S)+\sqrt{(A_S-C_S)^2+B_S^2} \right] \,, \\
m_{h}^2 &=& \frac{1}{2}\left[(A_S+C_S)-\sqrt{(A_S-C_S)^2+B_S^2} \right] \,. 
\end{eqnarray}
\end{subequations}
The physical scalars are obtained by rotating the original basis by an angle $\alpha$~:
\begin{eqnarray}
\begin{pmatrix} H \\ h \end{pmatrix} = \begin{pmatrix} \cos\alpha &
  \sin\alpha \\ 
-\sin\alpha & \cos\alpha \end{pmatrix} \begin{pmatrix} h_1 \\ h_2 \end{pmatrix} \,.
\end{eqnarray}
This rotation angle is defined through the following relation
\begin{eqnarray}
\tan 2\alpha =\frac{2B_S}{A_S-C_S} =
\frac{2\left(\lambda_3+\frac{\lambda_5}{4}
  \right)v_1v_2}{\lambda_1v_1^2-
  \lambda_2v_2^2+\left(\lambda_3+\frac{\lambda_5}{4}\right)(v_1^2-v_2^2)}
\,.
\end{eqnarray}
Note that there were eight parameters to start with: $v_1,~v_2$ and 6
lambdas. We trade $v_1$ and $v_2$ for $v$ and $\tan\beta$. All the
lambdas except $\lambda_5$ can be traded for 4 physical scalar masses
and $\alpha$. The relations between these two equivalent sets of
parameters are given below~:
\begin{subequations}
\begin{eqnarray}
\lambda_1 &=& \frac{1}{2v^2\cos^2\beta}\left[m_H^2\cos^2\alpha
  +m_h^2\sin^2\alpha
  -\frac{\sin\alpha\cos\alpha}{\tan\beta}\left(m_H^2-m_h^2\right)\right]
-\frac{\lambda_5}{4}\left(\tan^2\beta-1\right) \,, \\
\lambda_2 &=& \frac{1}{2v^2\sin^2\beta}\left[m_h^2\cos^2\alpha
  +m_H^2\sin^2\alpha
  -\sin\alpha\cos\alpha\tan\beta\left(m_H^2-m_h^2\right) \right]
-\frac{\lambda_5}{4}\left(\cot^2\beta-1\right) \,, \\
\lambda_3 &=& \frac{1}{2v^2}
\frac{\sin\alpha\cos\alpha}{\sin\beta\cos\beta}
\left(m_H^2-m_h^2\right) -\frac{\lambda_5}{4} \,, \\
\lambda_4 &=& \frac{2}{v^2} m_{H^+}^2 \,, \\
\lambda_6 &=& \frac{2}{v^2} m_A^2 \,.
\end{eqnarray}
\label{inv2HDM}
\end{subequations}
Among these, $v$ is already known (246 GeV) and if we assume that the
lightest CP-even Higgs is what has been observed at the LHC, then
$m_h$ is also known (125 GeV). The rest of the parameters need to be
constrained from theoretical as well as experimental considerations.

\subsection{The alignment limit}
The {\em alignment limit} corresponds to recovering a CP-even scalar
mass eigenstate with exactly the same gauge, Yukawa and self couplings
couplings at tree level as those of the SM Higgs bosons.  To start
with, it is instructive to look at the trilinear gauge-Higgs couplings
which stem from the Higgs kinetic terms:
\begin{eqnarray}
{\mathscr L}_{\rm kin}^{\rm scalar} = |D_\mu\Phi_1|^2+|D_\mu\Phi_2|^2
\ni \frac{g^2}{2}W_\mu^+W^{\mu -}(v_1h_1+v_2h_2) \,.
\end{eqnarray}
Clearly, the combination
\begin{eqnarray}
H^0=\frac{1}{v}(v_1h_1+v_2h_2)
\end{eqnarray}
will have gauge couplings exactly as the SM Higgs boson and its
orthogonal combination ($R$) will not have any $RZZ$ or $RWW$
trilinear couplings.  $H^0$ also mimics the SM Higgs in Yukawa
couplings.  The states $H^0$ and $R$ can be obtained by applying the
same rotation as in the charged and pseudoscalar sectors:
\begin{eqnarray}
\begin{pmatrix} H^0 \\ R \end{pmatrix} = \begin{pmatrix} \cos\beta &
  \sin\beta \\ 
-\sin\beta & \cos\beta \end{pmatrix} \begin{pmatrix} h_1 \\ h_2 \end{pmatrix} \,.
\end{eqnarray}
Note that this SM-like state $H^0$ is not guaranteed to be a mass
eigenstate in general. The alignment limit specifically implies the
condition under which $H^0$ coincides with one of the CP-even physical
eigenstates. To go to the physical basis $(H,~h)$ from $(H^0,~R)$ one
needs the following rotation:
 \begin{subequations}
 \label{align_def}
 \begin{eqnarray}
 H &=& \cos(\beta-\alpha) H^0-\sin(\beta-\alpha) R \,, \\
 h &=& \sin(\beta-\alpha) H^0+\cos(\beta-\alpha)R \,.
 \end{eqnarray}
 \end{subequations}
Clearly, if we want the lightest CP-even scalar $h$ to posses SM-like
couplings, we must set $\sin(\beta-\alpha)=1$, which is the definition
of the alignment limit. Thus by going to this limit one more parameter
is reduced.

\begin{figure}
\includegraphics[scale=0.3]{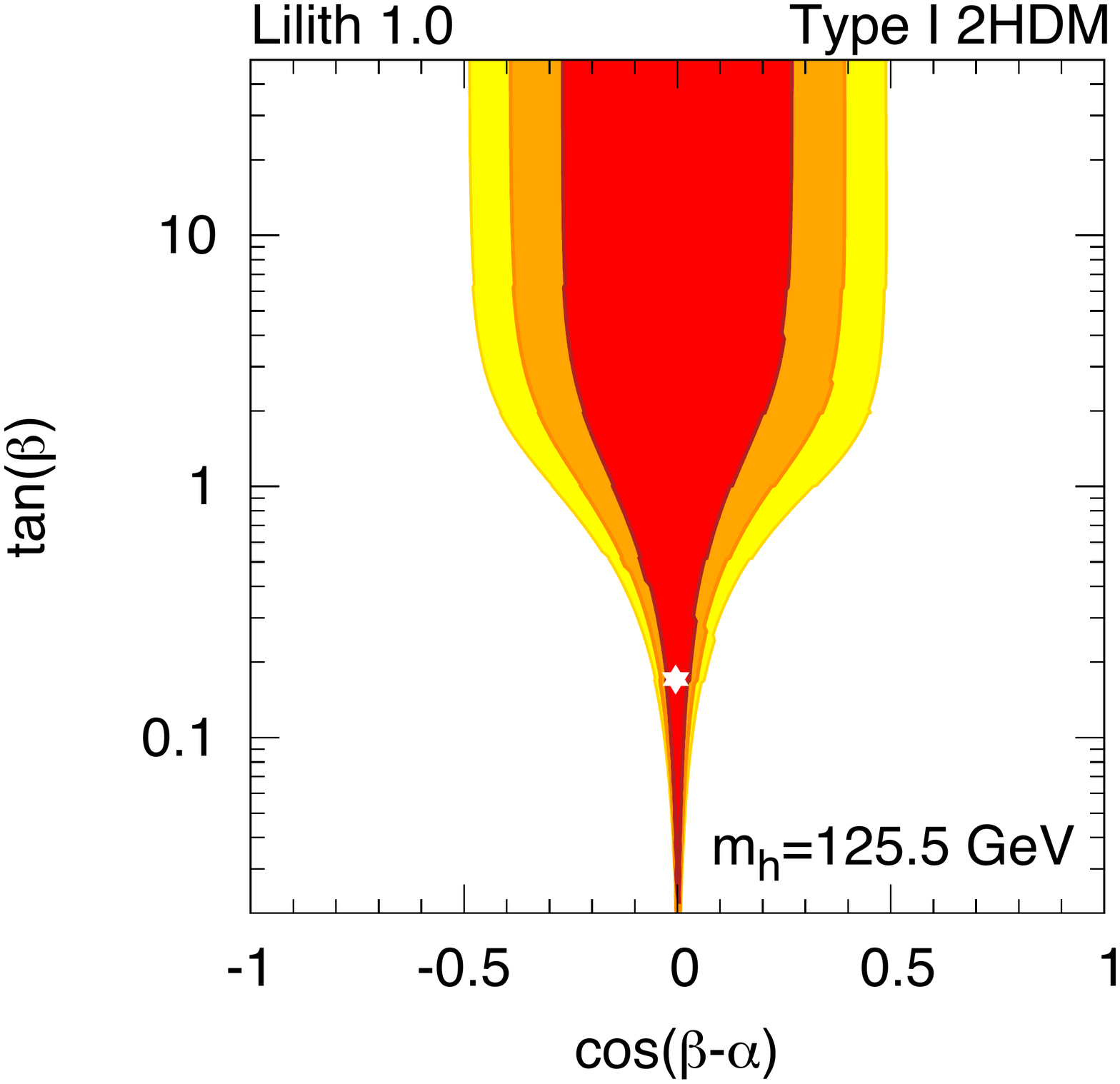} 
\includegraphics[scale=0.3]{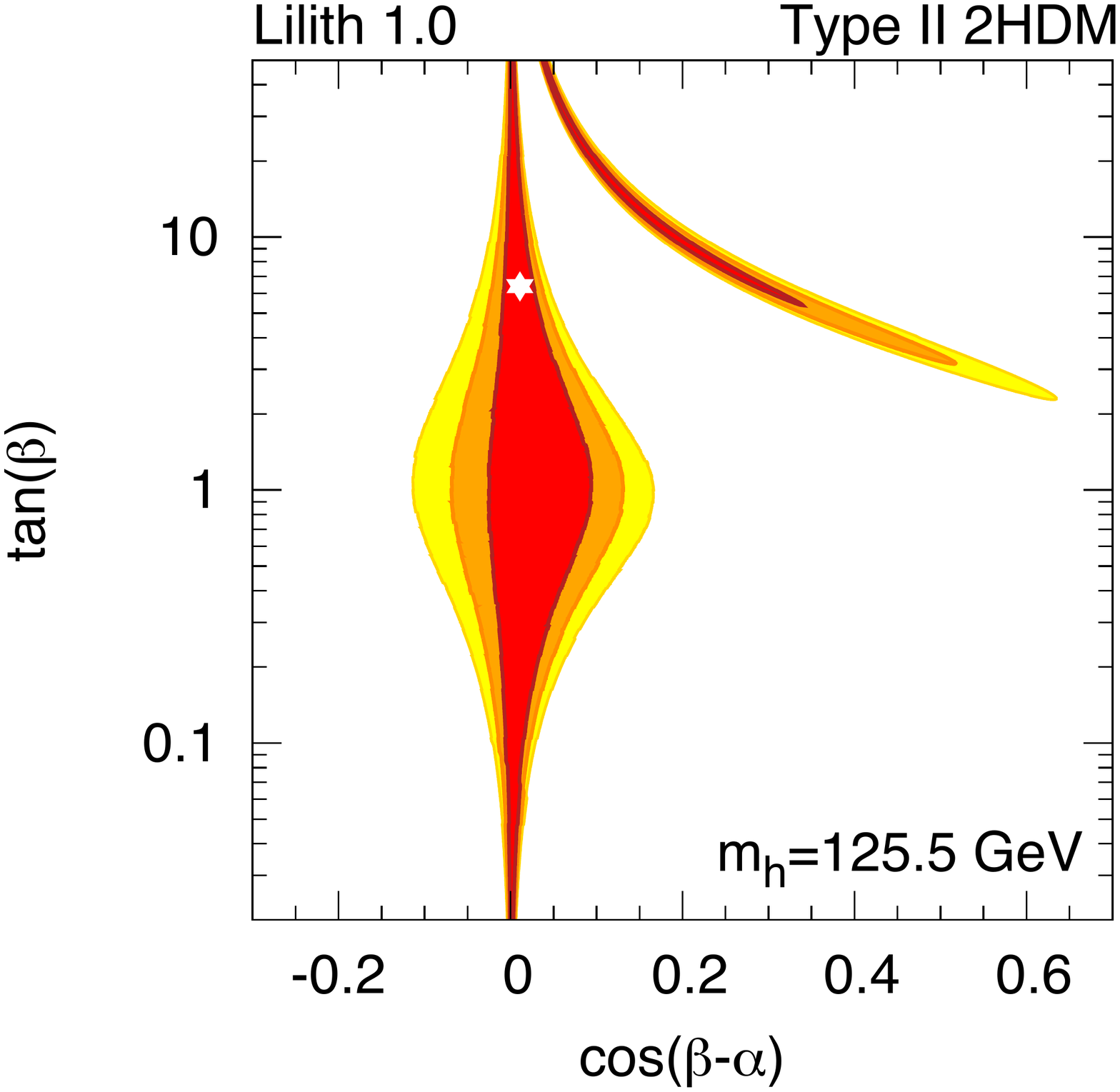}
\caption{\em The red, orange and yellow regions represent the 68\%,
  95\% and 99\% C.L. allowed regions, respectively, coming from the
  Higgs signal strength measurements at the LHC. The left (right)
  panel shows the situation for Type I (II) model. The star-marked
  points correspond to the best fit values (figures adapted from
  \cite{Bernon:2014vta}).}
\label{current_fit}
\end{figure}

\begin{figure}
\includegraphics[scale=0.36]{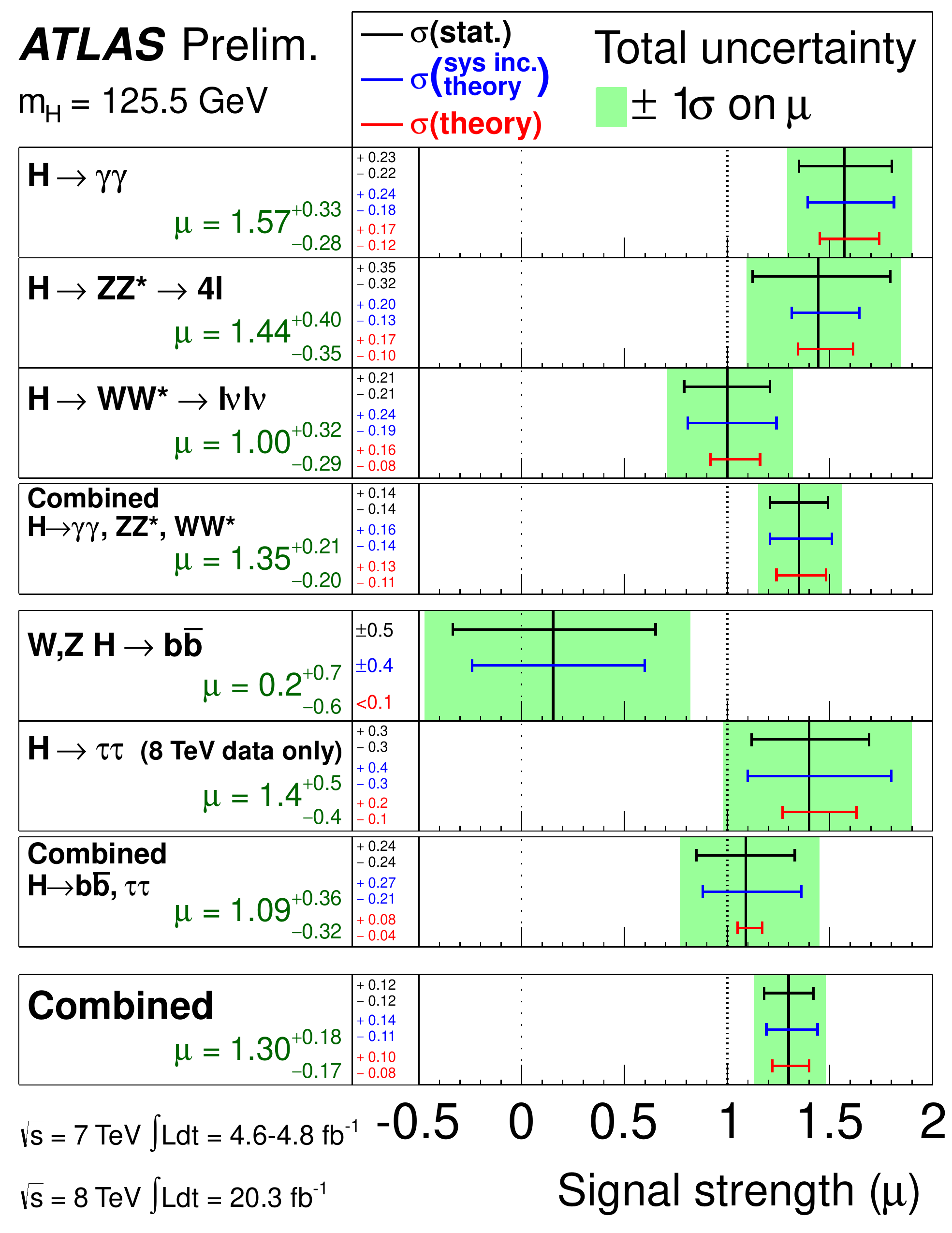} ~
\includegraphics[scale=0.45]{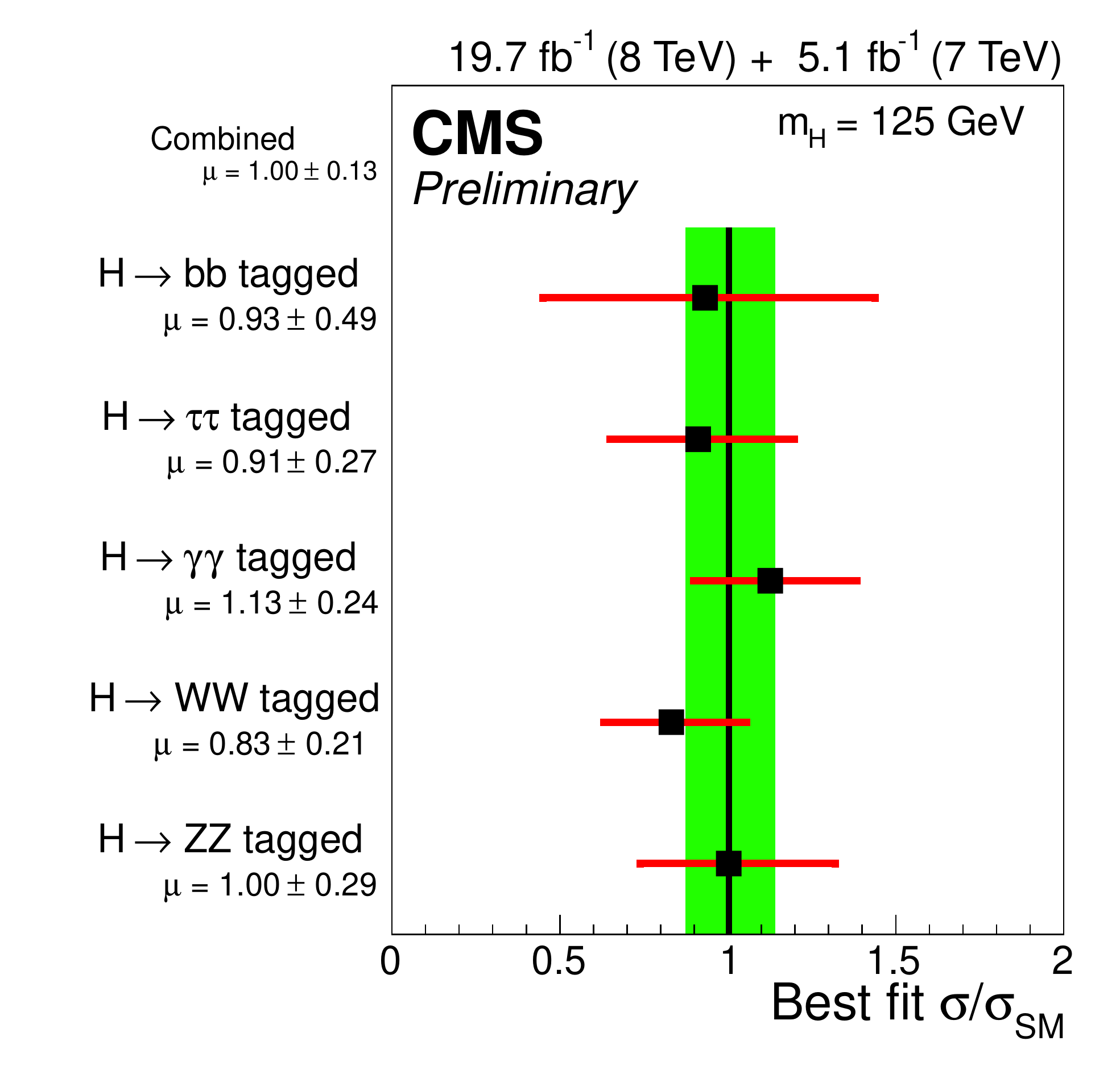}
\caption{\em Current measurements of the Higgs signal strengths into
  different channels by the ATLAS \cite{Atlas:signal} and CMS
  \cite{cms:signal} Collaborations.}
\label{signal_strengths}
\end{figure}

Now we come to the important question of how crucial the alignment
limit is in the context of current LHC Higgs data. Many global fit
results in view of the recent data can be found in the
literature\cite{Coleppa:2013dya,Eberhardt:2013uba, Chen:2013rba,
  Craig:2013hca, Dumont:2014wha, Bernon:2014vta}. In
Fig.~\ref{current_fit} we display the results of a recent
analysis\cite{Bernon:2014vta}. The orange part represents the 95\%
C.L. allowed region from measurements of the Higgs signal strengths in
various channels (for the latter, see
Fig.~\ref{signal_strengths}). Since the data is compatible with the SM
predictions, the alignment limit is preferred. The horizontal widths
of the allowed regions reflect the present accuracy of
measurements. In Ref.~\cite{Dumont:2014wha}, it has been shown how
this width will shrink if future measurements continue to agree more
with the SM predictions with greater accuracy, thus pushing us closer
to the alignment limit. The SM alignment limit has also been motivated
by employing different global symmetries of the scalar potential
\cite{Dev:2015bta}.

%% file: Stability.tex
\section{Stability constraints}
To discuss the stability of the potential it is convenient to work
with the notation of \Eqn{notation1}. Here we derive the constraints
on parameters $\beta_i$ such that the scalar potential $V$ is bounded
from below in any direction in the field space\cite{Deshpande:1977rw,Kastening:1992by}. It is sufficient to
examine the quartic terms of the scalar potential (which we denote by
$V_4$) because only this part of the potential will be dominant for
large values of the field components of $\Phi_1$ and $\Phi_2$. We
define $a \equiv \Phi_1^\dagger\Phi_1$, $b \equiv
\Phi_2^\dagger\Phi_2$, $c \equiv {\rm Re}~\Phi_1^\dagger\Phi_2$, $d
\equiv {\rm Im}~\Phi_1^\dagger\Phi_2$ and note that
\begin{eqnarray}
ab \ge c^2+d^2 \,.
\end{eqnarray}
Using these definitions we can rewrite the quartic part of the scalar
potential as follows~\cite{Gunion:2002zf}:
\begin{eqnarray}
V_4 &=& \frac{1}{2}\left(\sqrt{\beta_1}a-\sqrt{\beta_2}b \right)^2+
\left(\beta_3+\sqrt{\beta_1\beta_2}\right) \left(ab-c^2-d^2\right)
+2\left(\beta_3+\beta_4+\sqrt{\beta_1\beta_2}\right)c^2 \nonumber
\\ && +\left({\rm Re}~\beta_5-\beta_3-\beta_4-\sqrt{\beta_1\beta_2}
\right)\left(c^2-d^2\right) -2cd~ {\rm Im}~\beta_5 \,.
\label{quartic_stab}
\end{eqnarray}
Although we assume all the potential parameters to be real for our
phenomenological studies, our arguments on stability do not depend on
whether $\beta_5$ is real or complex.  We must ensure that $V_4$ never
becomes infinitely negative in any direction of the field space, {\it
  i.e.}, for any choice of 8 independent field parameters (4 of
$\Phi_1$ and 4 of $\Phi_2$). Since $\Phi_1$ and $\Phi_2$ are two
component column matrices, it is possible to choose arbitrary nonzero
values for $a$ and $b$ even when we make $c=d=0$. But if $a$ and/or
$b$ becomes zero, then $c=d=0$ automatically. Keeping these in mind,
we now proceed to find the stability constraints, i.e. the conditions
under which the potential is bounded from below\cite{Deshpande:1977rw}.
\begin{enumerate}[($i$)]
\item Consider the field direction $b=0$ (and therefore $c=d=0$) and
  $a\to\infty$; then $V_4=(\beta_1/2)a^2$. So, $V_4$ is not largely
  negative requires
\begin{equation}
\beta_1 \ge 0\,. \label{stab1}
\end{equation}
\item Consider the field direction $a=0$ (and therefore $c=d=0$) and
  $b\to\infty$; then $V_4=(\beta_2/2)b^2$. So, $V_4$ is not largely
  negative requires
\begin{equation}
\beta_2 \ge 0 \,. \label{stab2}
\end{equation}
\item Consider the field direction along which
  $a=\sqrt{\beta_2/\beta_1}b$ (so that the first term in
  \Eqn{quartic_stab} vanishes) and $c=d=0$. In addition to this we go
  to large field values in that direction, {\it i.e.},
  $a,~b\to\infty$. Then, $V_4=(\beta_3+\sqrt{\beta_1\beta_2})ab$. Now,
  as $a,~b >0$ by definition, the condition for the potential not to
  hit $(-\infty)$ becomes
\begin{equation}
\beta_3+\sqrt{\beta_1\beta_2} \ge 0 \,. \label{stab3}
\end{equation}
\item Again consider the field direction in which
  $a=\sqrt{\beta_2/\beta_1}b$ along with $ab=c^2+d^2$. Along this
  direction, $V_4$ is of the form
 \begin{subequations}
 \begin{eqnarray}
 V_4 &=& Pc^2+2Qcd+Rd^2 \,, \label{v4_4th} \\
 {\rm where,~~} P &=& {\rm Re}~\beta_5+\Lambda \,, \\
 Q &=& -{\rm Im}~\beta_5 \,, \\
 R &=& -{\rm Re}~\beta_5+\Lambda \,, \\
 {\rm with,}~~ S &=& \beta_3+\beta_4+\sqrt{\beta_1\beta_2} \,.
 \end{eqnarray}
 \end{subequations}
  Since $c$ and $d$ are still arbitrary, by choosing
 $d=0,~c\to\infty$ and $c=0,~d\to\infty$ successively, we require
  \begin{subequations}
  \label{Lamb_pre}
  \begin{eqnarray}
  P &=& {\rm Re}~\beta_5+ S \ge 0 \,, \label{P_vac} \\
  R &=& -{\rm Re}~\beta_5+ S \ge 0 \,, \\
  {\rm and~ hence,} && S \ge 0 \,.
  \label{stab_aux}
  \end{eqnarray}
  \end{subequations}
\end{enumerate}
To have another condition, let us recast \Eqn{v4_4th} into the
following form~:
\begin{eqnarray}
V_4 &=& P\left(c+\frac{Q}{P}d\right)^2 + \left(R-\frac{Q^2}{P}\right)d^2 \,.
\label{v4recast}
\end{eqnarray}
We can now choose a direction along which $c=-Q/P~d$ with $d\to\infty$
so that we have the following condition~:
\begin{eqnarray}
R-\frac{Q^2}{P} > 0 ~~\Rightarrow PR>Q^2 \,.
\label{vac_final}
\end{eqnarray}
For the last step, remember that $P>0$ (\Eqn{P_vac}) so that we can
multiply both sides by $P$ without flipping the inequality sign. After
substituting for $P$, $Q$ and $R$ we get from \Eqn{vac_final}~:
 \begin{subequations}
 \begin{eqnarray}
&& S^2-\left({\rm Re}~\beta_5\right)^2 > \left({\rm
     Im}~\beta_5\right)^2 ~~\Rightarrow S^2 >|\beta_5|^2 \,, \\ && S >
   |\beta_5| \,,
 \label{Lamb}
 \end{eqnarray}
 \end{subequations}
 where, in the last step we have used the fact that $S>0$
 (\Eqn{stab_aux}). Since $|\beta_5|>\pm\beta_5,~0$, \Eqn{Lamb} puts a
 stronger constrain on $S$ than \Eqn{Lamb_pre}. Therefore,
 substituting for $S$, \Eqn{Lamb} becomes
 \begin{eqnarray}
 \beta_3+\beta_4+\sqrt{\beta_1\beta_2} > |\beta_5| \,.
 \label{stab4}
 \end{eqnarray}
 We now collect Eqs.~(\ref{stab1}), (\ref{stab2}), (\ref{stab3}) and
 (\ref{stab4}) together and, using \Eqn{connections}, express them in
 terms of lambdas for later use~:
  \begin{subequations}
  \label{stability}
  \begin{eqnarray}
&& \lambda_1+\lambda_3 >0 \,, \\ && \lambda_2+\lambda_3 >0 \,, \\ &&
    (2\lambda_3+\lambda_4)
    +2\sqrt{(\lambda_1+\lambda_3)(\lambda_2+\lambda_3)} >0 \,, \\ &&
    2\lambda_3+\frac{\lambda_5+\lambda_6}{2}
    -\frac{|\lambda_5-\lambda_6|}{2}+2\sqrt{(\lambda_1+\lambda_3)(\lambda_2+\lambda_3)}
    >0 \,.
  \end{eqnarray}
  \end{subequations}
The above conditions are both necessary and sufficient for ensuring
the stability of the electroweak vacuum. This has also been shown
through rigorous analysis in \cite{Klimenko:1984qx, Maniatis:2006fs}.
In addition to these stability conditions, one also needs to ensure
the positivity of the physical scalar
masses\cite{Kastening:1992by,Nie:1998yn}. Additionally, one might also
wish to ensure that the minima is indeed the global minimum.  The
condition for the latter is given by\cite{Barroso:2013awa}
\begin{eqnarray}
m_{12}^2\left(m_{11}^2-m_{22}^2\sqrt{\frac{\beta_1}{\beta_2}}\right)
\left(\tan\beta -\sqrt[4]{\frac{\beta_1}{\beta_2}}\right) > 0 \,.
\end{eqnarray}

%% file: Unitarity.tex
\section{Unitarity constraints}
In this section we study the energy growth of scattering amplitudes
involving the scalar states.  Any scattering amplitude can be expanded
in terms of the partial waves as follows~:
\begin{equation}
\mathcal{M}(\theta) = 16 \pi\sum\limits_{\ell=0}^{\infty} a_\ell
(2\ell+1)P_\ell (\cos\theta) \,,
\label{def:feynman}
\end{equation}
where, $\theta$ is the scattering angle and $P_\ell (x)$ is the
Legendre polynomial of order $\ell$. The prescription is as follows:
once we calculate the Feynman amplitude of a certain $2\to 2$
scattering process, each of the partial wave amplitude ($a_\ell$), in
\Eqn{def:feynman}, can be extracted by using the orthonormality of the
Legendre polynomials.  In the context of SM, the pioneering work has
been done by Lee, Quigg and Thacker (LQT) \cite{Lee:1977eg}. They have
analyzed several two body scatterings involving longitudinal gauge
bosons and physical Higgs in the SM. All such scattering amplitudes
are proportional to Higgs quartic coupling in the high energy
limit. The $\ell=0$ partial wave amplitude $(a_0)$ is then extracted
from these amplitudes and cast in the form of what is called an
S-matrix having different two-body states as rows and columns. The
largest eigenvalue of this matrix is bounded by the unitarity
constraint, $|a_0 | < 1$.  This restricts the quartic Higgs self
coupling and therefore the Higgs mass to a maximum value.

The procedure has been extended to the case of a 2HDM scalar potential
\cite{Maalampi:1991fb,Kanemura:1993hm,Akeroyd:2000wc,Horejsi:2005da,Swiezewska:2012ej}.
Here also same types of two body scattering channels are
considered. Thanks to the equivalence theorem
\cite{Pal:1994jk,Horejsi:1995jj}, we can use unphysical Higgses
instead of actual longitudinal components of the gauge bosons when
considering the high energy limit. The diagrams containing trilinear
vertices will be suppressed by a factor of $E^2$ coming from the
intermediate propagator. Thus they do not contribute at high energies,
and only the quartic couplings contribute.  Since we are interested
only in the eigenvalues of the S-matrix, we may for convenience with
the original fields of \Eqn{notation2} instead of the physical mass
eigenstates.

To provide clarity, let us outline the method of obtaining the
constraints.  As already argued, only the dimensionless quartic
couplings will contribute to the amplitudes at high energies. As a
result, only $\ell=0$ partial amplitude ($a_0$) will receive nonzero
contribution from the leading order terms in the scattering
amplitudes. The task is to find the expressions of $a_0$ for every
possible $2\to 2$ scattering process and cast them in the form of an
S-matrix which is constructed by taking the different two-body
channels as rows and columns. Unitarity will restrict the magnitude of
each of the eigenvalues of this S-matrix to lie below unity.

First we identify all the possible two-particle channels. These
two-particle states are made of the fields $w_{k}^{\pm},~h_{k}$ and
$z_{k}$ corresponding to the parametrization of \Eqn{phi}. For our
calculation, we consider neutral combinations out of two-particle
states (e.g., $w_{i}^{+}w_{j}^{-},~h_i h_j,~z_i z_j,~h_i z_j$) and
singly charged two-particle states (e.g.,
$w_{i}^{+}h_j,~w_{i}^{+}z_j$). In general, if we have $n$-number of
doublets $\phi_{k}~(k=1,\ldots,n)$ in $n$HDM scenarios, there will be
$(3n^2+n)$-number of neutral and $2n^2$-number of charged two-particle
states. Clearly, the dimensions of S-matries formed out of these
two-particle states will be a $(3n^2+n)\times(3n^2+n)$ and $2n^2\times
2n^2$ for the neutral and charged cases respectively. The eigenvalues
of these matrices should be bounded by the unitarity constraint.

The neutral channel S-matrix for 2HDM is a $14\times 14$ matrix with the
following two-particle states as rows and columns~:
\begin{eqnarray}
w_1^+w_1^-,~
w_2^+w_2^-,~w_1^+w_2^-,~w_2^+w_1^-,~\frac{h_1h_1}{\sqrt{2}},~\frac{z_1z_1}{\sqrt{2}},
~\frac{h_2h_2}{\sqrt{2}},~\frac{z_2z_2}{\sqrt{2}},~
h_1z_2,~h_2z_1,~z_1z_2,~h_1h_2,~h_1z_1,~h_2z_2\,.
\end{eqnarray}
The factor of $1/\sqrt{2}$ associated with the identical particle
states arises due to Bose symmetry. In the most general case, finding
the eigenvalues of the $14\times 14$ matrix would be a tedious
job. But the potential of \Eqn{notation2} contains some obvious
symmetries in its quartic terms. These symmetries will allow us to
decompose the full matrix in smaller blocks. Now, each term in the
quartic part of the potential always contains even number of indices
(1 or 2). Consequently a state $x_1y_1$ or $x_2y_2$ will always
scatter into $x_1y_1$ or $x_2y_2$ but not into $x_1y_2$ or $x_2y_1$
and vice versa. Furthermore, CP symmetry is conserved. This implies
that a neutral combination $h_ih_j$ or $z_iz_j$ will never go into
$h_iz_j$. Keeping these facts in mind we can now decompose the
S-matrix in the neutral sector into smaller blocks as follows~:
\begin{equation}
\mathcal{M}_{N}= \begin{pmatrix} (\mathcal{M}_N^{11})_{6\times 6} & 0
  & 0 \\ 0 & (\mathcal{M}_N^{11})_{2\times 2} & 0 \\ 0 & 0 &
  (\mathcal{M}_N^{12})_{6\times 6}
\end{pmatrix} \,.
\label{mat_neutral}
\end{equation} 
The submatrices are given below~:
\small
\begin{subequations}
\begin{eqnarray}
 (\mathcal{M}_N^{11})_{6\times 6} =
  \bordermatrix{
  &\m w_1^+ w_1^-&\m  w_2^+ w_2^-&\m \frac{z_1z_1}{\sqrt{2}} &\m \frac{h_1h_1}{\sqrt{2}}&\m \frac{z_2z_2}{\sqrt{2}}&\m \frac{h_2h_2}{\sqrt{2}} \cr\vbox{\hrule}
\m w_1^+ w_1^- & 4(\l_1+\l_3) & 2\l_3+\frac{\l_5+\l_6}{2}  & \sqrt{2}(\l_1+\l_3) & \sqrt{2}(\l_1+\l_3) & \sqrt{2}\left(\l_3+\frac{\l_4}{2}\right)  & \sqrt{2}\left(\l_3+\frac{\l_4}{2}\right)
   \cr
\m w_2^+ w_2^-& 2\l_3+\frac{\l_5+\l_6}{2} & 4(\l_2+\l_3) & \sqrt{2}\left(\l_3+\frac{\l_4}{2}\right) & \sqrt{2}\left(\l_3+\frac{\l_4}{2}\right) &  \sqrt{2}(\l_2+\l_3) & \sqrt{2}(\l_2+\l_3)
   \cr
\m \frac{z_1z_1}{\sqrt{2}} &  \sqrt{2}(\l_1+\l_3) & \sqrt{2}\left(\l_3+\frac{\l_4}{2}\right) & 3(\l_1+\l_3) & (\l_1+\l_3) & \l_3+\frac{\l_5}{2} & \l_3+\frac{\l_6}{2}
   \cr
\m \frac{h_1h_1}{\sqrt{2}} &  \sqrt{2}(\l_1+\l_3) & \sqrt{2}\left(\l_3+\frac{\l_4}{2}\right) & (\l_1+\l_3) & 3(\l_1+\l_3) & \l_3+\frac{\l_6}{2} & \l_3+\frac{\l_5}{2} 
   \cr 
\m \frac{z_2z_2}{\sqrt{2}} &\sqrt{2}\left(\l_3+\frac{\l_4}{2}\right) &\sqrt{2}(\l_2+\l_3) & \l_3+\frac{\l_5}{2}& \l_3+\frac{\l_6}{2} & 3(\l_2+\l_3)  & (\l_2+\l_3)  
   \cr
\m \frac{h_2h_2}{\sqrt{2}} & \sqrt{2}\left(\l_3+\frac{\l_4}{2}\right) & \sqrt{2}(\l_2+\l_3)  &\l_3+\frac{\l_6}{2} & \l_3+\frac{\l_5}{2} & (\l_2+\l_3)  & 3(\l_2+\l_3)  \cr
   }\,, 
\end{eqnarray}
\begin{eqnarray}
 (\mathcal{M}_N^{11})_{2\times 2} =
  \bordermatrix{
  &\m h_1z_1 &\m  h_2z_2  \cr\vbox{\hrule}
\m h_1z_1 & 2(\l_1+\l_3) & \frac{\l_5-\l_6}{2}
   \cr
\m h_2z_2 & \frac{\l_5-\l_6}{2} & 2(\l_2+\l_3)
   \cr
   }\,, 
\end{eqnarray}
\begin{eqnarray}
 (\mathcal{M}_N^{12})_{6\times 6} =
  \bordermatrix{
  &\m w_1^+ w_2^-&\m  w_2^+ w_1^-&\m h_1z_2 &\m h_2z_1 &\m z_1z_2 &\m h_1h_2 \cr\vbox{\hrule}
\m w_1^+ w_2^- & 2\l_3+\frac{\l_5+\l_6}{2} & \l_5-\l_6  & -\frac{i}{2}(\l_4-\l_6) & \frac{i}{2}(\l_4-\l_6) & \frac{\l_5-\l_4}{2} & \frac{\l_5-\l_4}{2}
   \cr
\m w_2^+ w_1^-& \l_5-\l_6 & 2\l_3+\frac{\l_5+\l_6}{2} & \frac{i}{2}(\l_4-\l_6) &-\frac{i}{2}(\l_4-\l_6)  & \frac{\l_5-\l_4}{2} &  \frac{\l_5-\l_4}{2}
   \cr
\m h_1z_2 & \frac{i}{2}(\l_4-\l_6)  & -\frac{i}{2}(\l_4-\l_6)  & 2\l_3+\l_6 & \frac{\l_5-\l_6}{2} & 0 & 0
   \cr
\m h_2z_1 &  -\frac{i}{2}(\l_4-\l_6)  & \frac{i}{2}(\l_4-\l_6)  & \frac{\l_5-\l_6}{2} & 2\l_3+\l_6  & 0 & 0
   \cr 
\m z_1z_2 & \frac{\l_5-\l_4}{2} & \frac{\l_5-\l_4}{2} & 0 & 0 & 2\l_3+\l_5  & \frac{\l_5-\l_6}{2}  
   \cr
\m h_1h_2 & \frac{\l_5-\l_4}{2} & \frac{\l_5-\l_4}{2}  & 0 & 0 & \frac{\l_5-\l_6}{2}  & 2\l_3+\l_5  \cr
   }\,.
\end{eqnarray}
\end{subequations}
\normalsize
The same exercise can be repeated for the charged two-particle state
combinations. With the singly charged state combinations, it will be a
$8\times 8$ matrix which will take the following block diagonal form~:
\begin{equation}
\mathcal{M}_{C}= \begin{pmatrix}
(\mathcal{M}_C^{11})_{4\times 4} & 0 \\ 0 & (\mathcal{M}_C^{12})_{4\times 4} 
\end{pmatrix} \,.
\label{mat_charged}
\end{equation} 
The submatrices are given below~:
 \begin{subequations}
 \begin{eqnarray}
  (\mathcal{M}_C^{11})_{4\times 4} &=&
   \bordermatrix{
   &\m h_1w_1^+ &\m  h_2w_2^+ &\m z_1w_1^+ &\m z_2w_2^+  \cr\vbox{\hrule}
 \m h_1w_1^+ & 2(\l_1+\l_3) & \frac{\l_5-\l_4}{2} & 0 & -\frac{i}{2}(\l_4-\l_6)
    \cr
 \m h_2w_2^+ & \frac{\l_5-\l_4}{2} & 2(\l_2+\l_3) & -\frac{i}{2}(\l_4-\l_6) & 0
    \cr
 \m z_1w_1^+ & 0 & \frac{i}{2}(\l_4-\l_6) & 2(\l_1+\l_3) & \frac{\l_5-\l_4}{2} 
  \cr 
 \m z_2w_2^+  & \frac{i}{2}(\l_4-\l_6)  & 0 & \frac{\l_5-\l_4}{2} & 2(\l_2+\l_3) \cr 
    }\,,  \\
  (\mathcal{M}_C^{12})_{4\times 4} &=&
   \bordermatrix{
   &\m h_1w_2^+ &\m  h_2w_1^+ &\m z_1w_2^+ &\m z_2w_1^+  \cr\vbox{\hrule}
 \m h_1w_2^+ & 2\l_3+\l_4 & \frac{\l_5-\l_4}{2} & 0 & \frac{i}{2}(\l_4-\l_6)
    \cr
 \m h_2w_1^+ & \frac{\l_5-\l_4}{2} & 2\l_3+\l_4 & \frac{i}{2}(\l_4-\l_6) & 0
    \cr
 \m z_1w_2^+ & 0 & -\frac{i}{2}(\l_4-\l_6) & 2\l_3+\l_4 & \frac{\l_5-\l_4}{2} 
  \cr 
 \m z_2w_1^+  & -\frac{i}{2}(\l_4-\l_6)  & 0 & \frac{\l_5-\l_4}{2} & 2\l_3+\l_4 \cr 
    }\,.
 \end{eqnarray}
 \end{subequations}
The eigenvalues for these matrices are given by
\begin{itemize}
\item $({\mathcal{M}_N^{11}})_{6\times 6}$~: $a_1^\pm,~a_2^\pm,~a_3^\pm$.
\item $({\mathcal{M}_N^{11}})_{2\times 2}$~: $a_3^\pm$.
\item $({\mathcal{M}_N^{12}})_{6\times 6}$~: $b_1,~b_2,~b_3,~b_4,~b_5$,
  with $b_5$ twofold degenerate.
\item $({\mathcal{M}_C^{11}})_{4\times 4}$~: $a_2^\pm,~a_3^\pm$.
\item $({\mathcal{M}_C^{12}})_{4\times 4}$~: $b_2,~b_4,~b_5,~b_6$.
\end{itemize}
We also display the explicit expressions for these eigenvalues~:
 \begin{subequations}
 \begin{eqnarray}
 a_1^\pm &=& 3(\l_1+\l_2+2\l_3) \pm \sqrt{9(\l_1-\l_2)^2+
   \left(4\l_3+\l_4+\frac{\l_5+\l_6}{2} \right)^2} \,, \\ a_2^\pm &=&
 (\l_1+\l_2+2\l_3) \pm \sqrt{(\l_1-\l_2)^2+\frac{1}{4}
   \left(2\l_4-\l_5-\l_6 \right)^2} \,, \\ a_3^\pm &=&
 (\l_1+\l_2+2\l_3) \pm \sqrt{(\l_1-\l_2)^2+\frac{1}{4} \left(\l_5-\l_6
   \right)^2} \,, \\ b_1 &=&
 2\l_3-\l_4-\frac{1}{2}\l_5+\frac{5}{2}\l_6 \,, \\ b_2 &=&
 2\l_3+\l_4-\frac{1}{2}\l_5+\frac{1}{2}\l_6 \,, \\ b_3 &=&
 2\l_3-\l_4+\frac{5}{2}\l_5-\frac{1}{2}\l_6 \,, \\ b_4 &=&
 2\l_3+\l_4+\frac{1}{2}\l_5-\frac{1}{2}\l_6 \,, \\ b_5 &=&
 2\l_3+\frac{1}{2}\l_5+\frac{1}{2}\l_6 \,, \\ b_6 &=&
 2(\l_3+\l_4)-\frac{1}{2}\l_5-\frac{1}{2}\l_6 \,.
 \end{eqnarray}
 \label{unieigenvalues}
 \end{subequations}
Each of the above eigenvalues will be bounded from the unitarity constraint as 
\begin{eqnarray}
|a_i^\pm|,~|b_i| \le 16\pi \,.
\end{eqnarray}

%% file: Combined_constraints.tex
\section{Numerical constraints on the scalar masses}
We now investigate the implications of the above conditions on the
physical scalar masses, especially the nonstandard ones.
\begin{figure}
\includegraphics[scale=1]{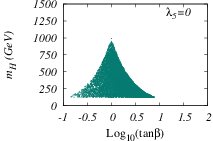}~~
\includegraphics[scale=1]{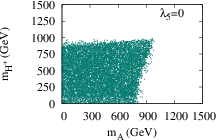}
\caption{\em Allowed region (shown by scattered points) from unitarity
  and stability for exact $Z_2$ symmetry ($\lambda_5=0$).}
\label{f:uniZ2}
\end{figure}
Fig.~\ref{f:uniZ2} shows the region allowed by the combined
constraints from unitarity and the boundedness of the potential for
$\lambda_5=0$, {\it i.e.}, exact $Z_2$ symmetry. Two noteworthy
features emerge \cite{Das:2015qva}~:
\begin{enumerate}[($i$)]
\item From the left panel of Fig.~\ref{f:uniZ2}, one can read the
  upper and lower limits on $\tan\beta$ as $1/8<\tan\beta<8$.
\item The upper limits on nonstandard scalars masses are given by:
  $m_H,~m_A,~m_{H^+}<1$ TeV.
\end{enumerate}
To understand the origin of the above limits we look into the
eigenvalues of \Eqn{unieigenvalues}. The first two constraints for
boundedness in \Eqn{stability} can be combined into
\begin{eqnarray}
\l_1+\l_2+2\l_3 > 0 \,.
\end{eqnarray}
Together with the condition $|a_1^\pm| < 16\pi$, this implies
\begin{eqnarray}
&& 0< \l_1+\l_2+2\l_3 < \frac{16\pi}{3} \,, \\ \Rightarrow && 0 <
  \left( m_H^2 -\frac{1}{2}\l_5v^2 \right)(\tan^2\beta+\cot^2\beta)
  +2m_h^2 <\frac{32\pi v^2}{3} \,,
\label{unitb}
\end{eqnarray}
where the last expression is obtained from the previous one by using
\Eqn{inv2HDM} in the alignment limit. Since the heavier CP-even Higgs
mass $m_H>125$ GeV, a limit on $\tan\beta$ (as well as $\cot\beta$) is
obtained when $\lambda_5=0$. Since the minimum value of
($\tan^2\beta+\cot^2\beta$) is 2 when $\tan\beta=1$, the maximum
possible value of $m_H$ is obtained for $\tan\beta=1$. \Eqn{unitb}
thus explains the $\tan\beta$ dependent bound on $m_H$ as depicted in
the left panel of Fig.~\ref{f:uniZ2}.

To obtain restrictions on the individual masses, $m_A$ and $m_{H^+}$,
we use the following inequalities~:
 \begin{subequations}
 \begin{eqnarray}
 |b_1-b_3| \equiv 3|\l_6-\l_5| < 32\pi  && \Rightarrow~
 |m_A^2-\frac{1}{2}\l_5v^2| < \frac{16\pi v^2}{3} \,, \label{limma}
 \\ b_6-b_3| \equiv 3|\l_4-\l_5| < 32\pi && \Rightarrow~
 |m_{H^+}^2-\frac{1}{2}\l_5v^2| < \frac{16\pi v^2}{3} \,.
 \label{limm1+}
 \end{eqnarray}
 \end{subequations}
Using Eqs.~(\ref{limma}) and (\ref{limm1+}) we put limits on $m_A$ and
$m_{H^+}$, respectively, when $\l_5=0$. Additionally, due to the
inequality
\begin{eqnarray}
|b_1-b_6| \equiv 3|\l_6-\l_4| < 32\pi && \Rightarrow~
|m_A^2-m_{H^+}^2| < \frac{16\pi v^2}{3} \,,
\label{diff}
\end{eqnarray}
we expect the splitting between $m_A$ and $m_{H^+}$ to be {\em always}
restricted. It is also interesting to note that the conclusions
obtained from Eqs.~(\ref{limma}), (\ref{limm1+}) and (\ref{diff}) do
{\em not} depend on the imposition of the alignment condition.

\begin{figure}
\includegraphics[scale=1]{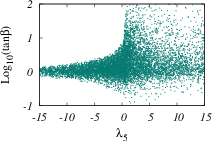}~~
\includegraphics[scale=1]{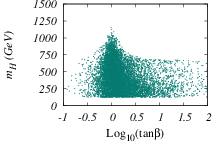}
\caption{\em Relaxation of the constraints on $\tan\beta$ for nonzero
  $\l_5$. The right panel should be compared with the left
  panel of Fig.~\ref{f:uniZ2}.}
\label{f:l5tb}
\end{figure}

From \Eqn{unitb} we also observe that the allowed space for
$\tan\beta$ is further squeezed if $\l_5 <0$, but the bound is relaxed
if $\l_5>0$. This feature emerges from the left panel of
Fig.~\ref{f:l5tb}. In addition to this, we can see from \Eqn{unitb}
that $m_H^2$ must be close to $1/2\l_5v^2$ if $\tan\beta$ moderately
deviates from unity. This feature is reflected by the horizontal tail
in the right panel of Fig.~\ref{f:l5tb} on both sides of the peak. The
width of the tail is a result of the variation of $\l_5$ in the range
[-15, 15].

\begin{figure}
\includegraphics[scale=0.66]{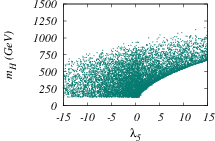}~~
\includegraphics[scale=0.66]{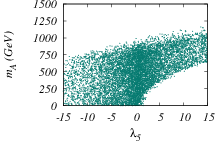}~~
\includegraphics[scale=0.66]{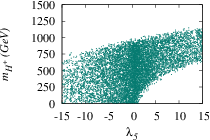}
\caption{\em Effect of nonzero $\l_5$ on the constraints on the nonstandard masses.}
\label{f:l5mass}
\end{figure}

From Eqs.~(\ref{limma}), (\ref{limm1+}) and (\ref{unitb}) we note that
the upper bounds on the nonstandard scalar masses will be relaxed for
$\l_5>0$ and get tighter for $\l_5<0$. Fig.~\ref{f:l5mass} reflects
these features.

\begin{figure}
\rotatebox{90}{\quad\quad\quad\quad$m_A$ (GeV)}
\includegraphics[scale=0.64]{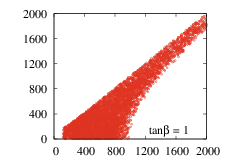} 
\includegraphics[scale=0.64]{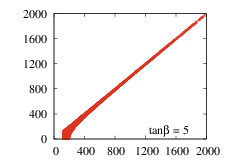}
\includegraphics[scale=0.64]{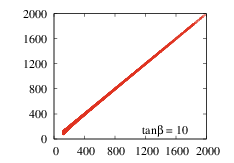}
\centerline{ \null \hfill $m_H$ (GeV) \quad}

\rotatebox{90}{\quad\quad\quad\quad$m_{H^+}$ (GeV)}
\includegraphics[scale=0.64]{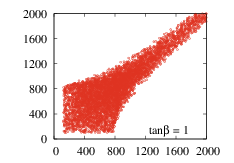} 
\includegraphics[scale=0.64]{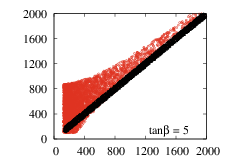}
\includegraphics[scale=0.64]{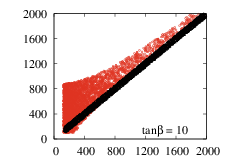}
\centerline{ \null \hfill $m_H$ (GeV) \quad}

\rotatebox{90}{\quad\quad\quad\quad$m_{H^+}$ (GeV)}
\includegraphics[scale=0.64]{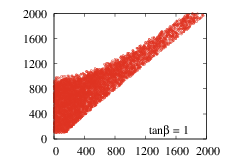} 
\includegraphics[scale=0.64]{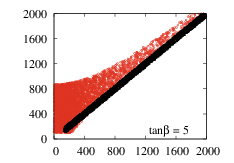}
\includegraphics[scale=0.64]{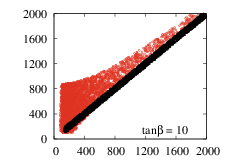}
\centerline{ \null \hfill $m_A$ (GeV) \quad}

\caption{\em 2HDM potential with softly broken U(1) symmetry: regions
  allowed in $m_H$-$m_A$, $m_H$-$m_{H^+}$ and $m_A$-$m_{H^+}$ planes
  from unitarity and stability (red points), and from $T$-parameter
  (black points), for three choices of $\tan\beta$.  The plots have
  been taken from \cite{Bhattacharyya:2013rya}, where $m_{H^+} >
  100$\,GeV was assumed to respect LEP direct search bound
  \cite{Searches:2001ac}.}
\label{f:uniu1}
\end{figure}

An interesting alternative arises if instead of $Z_2$ one imposes an
U(1) symmetry under which $\Phi_1 \to \Phi_1$ and $\Phi_2\to
e^{i\alpha}\Phi_2$. This U(1) symmetry needs to be softly broken to
forbid the appearance of an exactly massless pseudoscalar. This U(1)
symmetry, in the quartic terms, implies $\beta_5=0$ in \Eqn{notation1}
or $\l_5=\l_6$ in \Eqn{notation2}. The constraints on the scalar
masses imposed by the stability and unitarity conditions in
\Eqs{stability}{unieigenvalues} have been plotted in
Fig.~\ref{f:uniu1} for $\tan\beta=1$, 5 and 10 by performing random
scan over all nonstandard scalar masses.

The following salient features emerge from the plots.
\begin{enumerate}[($i$)]
\item There is a correlation between $m_A$ and $m_H$ which gets
  stronger for larger values of $\tan\beta$. They become nearly
  degenerate once $\tan\beta > 10$.  To understand this, we observe
  that \Eqn{unitb} for $\l_5=\l_6$ reduces to
\begin{eqnarray}
0 \leq (m_H^2-m_A^2)(\tan^2\beta+\cot^2\beta)+2m_h^2 
\leq {\frac{32\pi v^2}{3}} \,.
\label{uniM1}
\end{eqnarray}
Clearly, for $\tan\beta$ away from unity, $H$ and $A$ are nearly
degenerate.

\item There is a similar correlation between $m_H$ and $m_{H^+}$, but
  unlike the previous point, without any dependence on $\tan\beta$.
  This can again be seen from the inequalities of \Eqs{limma}{limm1+}
  keeping in mind that now $m_A^2=(1/2)\l_5v^2$.

\item The unitarity conditions essentially apply on the difference of
  the nonstandard squared masses.  Any individual mass can be
  arbitrarily large without affecting the unitarity conditions.  This
  conclusion crucially depends on the existence of a U(1) symmetry and
  its soft breaking term in the potential.  When the symmetry of the
  potential is only a discrete $Z_2$, considerations of unitarity do
  restrict the individual nonstandard masses as has already been
  shown.

\item The splitting between the heavy scalar masses is also
  constrained by the oblique electroweak $T$-parameter, whose
  expression in the decoupling limit is given by~\cite{He:2001tp,
    Grimus:2007if}
\begin{eqnarray}
T = {1 \over 16\pi \sin^2 \theta_w M_W^2} \Big[ F(m_{H^+}^2,m_H^2) +
  F(m_{H^+}^2, m_A^2) - F(m_H^2,m_A^2) \Big] \,,
\label{T}
\end{eqnarray}
with
\begin{eqnarray}
F(x,y) = {x+y \over 2} - {xy \over x-y} \; \ln (x/y) \,.
\end{eqnarray}
The new physics part in the $T$-parameter is given by ~\cite{Baak:2013ppa} 
\begin{eqnarray}
T = 0.05 \pm 0.12 \,.
\end{eqnarray}

To provide intuition into the constraints from the $T$-parameter, we
assume $m_H=m_A$, which is anyway dictated by the unitarity
constraints for $\tan\beta$ somewhat away from unity.  It then follows
from \Eqn{T} that the splitting between $m_{H^+}$ and $m_H$ is
approximately 50\,GeV, for $|m_{H^+}-m_H| \ll m_{H^+},m_H$.  It
follows from Fig.~\ref{f:uniu1} that the constraints from the
$T$-parameter are stronger than those from unitarity and stability.

On the other hand, for $\tan\beta=1$, unitarity and stability do not
compel $m_H$ and $m_A$ to be very close.  Then the $T$-parameter
cannot give any definitive constraints in the planes of the
nonstandard scalar masses, unlike the unitarity and stability
constraints.  For this reason, we have shown only the
unitarity/stability constraints in Fig.~\ref{f:uniu1} for
$\tan\beta=1$.

\item For moderate or large $\tan\beta$, the unitarity and stability
  constraints together with the $T$-parameter constraints imply that
  all three heavy scalar states are nearly degenerate in the alignment
  limit.
\end{enumerate}

For completeness, we also comment on the high scale validity of 2HDMs
\cite{Ferreira:2009jb}.  After the Higgs discovery, it is well known
that the SM perhaps needs to be augmented by new physics at energies
beyond $10^8$-$10^{10}$~GeV
\cite{Bezrukov:2012sa,Degrassi:2012ry}. The reason behind this is the
following: the Higgs boson mass $\sim 125$ GeV turns out to be a
little `smaller' than what could have made the SM ultraviolet safe up
to Planck scale. More precisely, in the tussle between the top-Yukawa
interaction (which tries to pull down the scalar quartic coupling) and
the quartic self interaction (which tries to push itself up) during
the course of RG evolution, the top-Yukawa takes an upper hand and
pulls the SM quartic coupling to negative values well below the Planck
scale.  One might expect the situation to improve in a 2HDM because of
the presence of additional quartic couplings. But the 2HDM potential
of \Eqs{notation1}{notation2}, without the soft breaking term, still
fails to maintain stability all the way up to the Planck scale
\cite{Chakrabarty:2014aya,Das:2015mwa,Chowdhury:2015yja}.  Even in the
presence of a soft breaking parameter, in the alignment limit, a lower
bound on $\tan\beta$ ($\gtrsim 3$)
\cite{Das:2015mwa,Chowdhury:2015yja} is obtained from the requirement
that the effect of the top-Yukawa ($= \sqrt{2} m_t/ (v\sin\beta)$ in
2HDM {\em vis-\`a-vis} $\sqrt{2} m_t/ v$ in SM) is sufficiently
diluted to maintain the required stability. Moreover, certain
correlations between the nonstandard masses and the soft breaking
parameter need to be maintained.

%% file: Conclusions.tex
\section{Conclusions and outlook}
Based on data on the scalar resonance observed at the LHC, two
important conclusions have been drawn: ($i$) its mass is approximately
125 GeV and ($ii$) its couplings to gauge bosons and fermions are
SM-like. In the 2HDM framework, the latter observation pushes us to
the {\em alignment limit}.  Put these two conditions together, the
2HDM parameter space is more constrained then ever.  Some important
observations in this context are the following:
\begin{itemize}

\item When the potential has a global U(1) symmetry, rather than a
  discrete $Z_2$, as well as a soft breaking term, unitarity restricts
  the mass-squared differences of the nonstandard scalars.  So the
  individual nonstandard masses, like $m_H, m_A$ or $m_{H^+}$, can
  grow very large without necessarily violating unitarity.

\item Strictly when both doublets receive vevs {\em and} the potential
  has an {\em exact} discrete or a global continuous symmetry, the
  model exhibits features of nondecoupling. For example, observables
  do not necessarily reduce to their SM values even when nonstandard
  scalars are too heavy. In \cite{Bhattacharyya:2014oka}, we have
  demonstrated this behavior in the context of $h \to \gamma\gamma$,
  where in the limit of an exact discrete symmetry, the Higgs signal
  strength in the diphoton channel at the LHC, given by
  $\mu_{\gamma\gamma}$, was shown to retain nonstandard effects even
  when the charged Higgs mass is pushed to extremely large
  values. Employing the soft breaking term, it is possible to ensure
  decoupling. If the symmetry to start with is a discrete $Z_2$, one
  cannot avoid fine-tuning between the charged Higgs mass and the soft
  breaking parameter to reach the decoupling limit. On the other hand,
  if the starting symmetry is a global U(1), the soft breaking
  parameter gets related to the pseudoscalar mass. In this case, the
  combined constraints from unitarity and the $T$-parameter naturally
  lead to the decoupling limit without any need for fine-tuning.  For
  details, we refer the readers to Ref.~\cite{Bhattacharyya:2014oka}.

\end{itemize}

\noindent{\bf Acknowledgments:}~ DD thanks Department of Atomic
Energy, India, for financial support.

%% file: main.bbl
\providecommand{\href}[2]{#2}\begingroup\raggedright\begin{thebibliography}{10}

\bibitem{Aad:2012tfa}
{\bf ATLAS} Collaboration, G.~Aad et~al., {\it {Observation of a new particle
  in the search for the Standard Model Higgs boson with the ATLAS detector at
  the LHC}},  {\em Phys.Lett.} {\bf B716} (2012) 1--29,
  [\href{http://arxiv.org/abs/1207.7214}{{\tt arXiv:1207.7214}}].

\bibitem{Chatrchyan:2012ufa}
{\bf CMS} Collaboration, S.~Chatrchyan et~al., {\it {Observation of a new boson
  at a mass of 125 GeV with the CMS experiment at the LHC}},  {\em Phys.Lett.}
  {\bf B716} (2012) 30--61, [\href{http://arxiv.org/abs/1207.7235}{{\tt
  arXiv:1207.7235}}].

\bibitem{Higgs:1964pj}
P.~W. Higgs, {\it {Broken Symmetries and the Masses of Gauge Bosons}},  {\em
  Phys.Rev.Lett.} {\bf 13} (1964) 508--509.

\bibitem{Higgs:1966ev}
P.~W. Higgs, {\it {Spontaneous Symmetry Breakdown without Massless Bosons}},
  {\em Phys.Rev.} {\bf 145} (1966) 1156--1163.

\bibitem{Englert:1964et}
F.~Englert and R.~Brout, {\it {Broken Symmetry and the Mass of Gauge Vector
  Mesons}},  {\em Phys.Rev.Lett.} {\bf 13} (1964) 321--323.

\bibitem{Guralnik:1964eu}
G.~Guralnik, C.~Hagen, and T.~Kibble, {\it {Global Conservation Laws and
  Massless Particles}},  {\em Phys.Rev.Lett.} {\bf 13} (1964) 585--587.

\bibitem{Kibble:1967sv}
T.~Kibble, {\it {Symmetry breaking in nonAbelian gauge theories}},  {\em
  Phys.Rev.} {\bf 155} (1967) 1554--1561.

\bibitem{Langacker:1980js}
P.~Langacker, {\it {Grand Unified Theories and Proton Decay}},  {\em
  Phys.Rept.} {\bf 72} (1981) 185.

\bibitem{Agashe:2014kda}
{\bf Particle Data Group} Collaboration, K.~Olive et~al., {\it {Review of
  Particle Physics}},  {\em Chin.Phys.} {\bf C38} (2014) 090001.

\bibitem{Branco:2011iw}
G.~Branco, P.~Ferreira, L.~Lavoura, M.~Rebelo, M.~Sher, et~al., {\it {Theory
  and phenomenology of two-Higgs-doublet models}},  {\em Phys.Rept.} {\bf 516}
  (2012) 1--102, [\href{http://arxiv.org/abs/1106.0034}{{\tt
  arXiv:1106.0034}}].

\bibitem{Dorsch:2013wja}
G.~Dorsch, S.~Huber, and J.~No, {\it {A strong electroweak phase transition in
  the 2HDM after LHC8}},  {\em JHEP} {\bf 1310} (2013) 029,
  [\href{http://arxiv.org/abs/1305.6610}{{\tt arXiv:1305.6610}}].

\bibitem{Glashow:1976nt}
S.~L. Glashow and S.~Weinberg, {\it {Natural Conservation Laws for Neutral
  Currents}},  {\em Phys.Rev.} {\bf D15} (1977) 1958.

\bibitem{Paschos:1976ay}
E.~Paschos, {\it {Diagonal Neutral Currents}},  {\em Phys.Rev.} {\bf D15}
  (1977) 1966.

\bibitem{Pich:2009sp}
A.~Pich and P.~Tuzon, {\it {Yukawa Alignment in the Two-Higgs-Doublet Model}},
  {\em Phys.Rev.} {\bf D80} (2009) 091702,
  [\href{http://arxiv.org/abs/0908.1554}{{\tt arXiv:0908.1554}}].

\bibitem{Ferreira:2010xe}
P.~Ferreira, L.~Lavoura, and J.~P. Silva, {\it {Renormalization-group
  constraints on Yukawa alignment in multi-Higgs-doublet models}},  {\em
  Phys.Lett.} {\bf B688} (2010) 341--344,
  [\href{http://arxiv.org/abs/1001.2561}{{\tt arXiv:1001.2561}}].

\bibitem{Branco:1996bq}
G.~Branco, W.~Grimus, and L.~Lavoura, {\it {Relating the scalar flavor changing
  neutral couplings to the CKM matrix}},  {\em Phys.Lett.} {\bf B380} (1996)
  119--126, [\href{http://arxiv.org/abs/hep-ph/9601383}{{\tt hep-ph/9601383}}].

\bibitem{Bhattacharyya:2014nja}
G.~Bhattacharyya, D.~Das, and A.~Kundu, {\it {Feasibility of light scalars in a
  class of two-Higgs-doublet models and their decay signatures}},  {\em
  Phys.Rev.} {\bf D89} (2014), no.~9 095029,
  [\href{http://arxiv.org/abs/1402.0364}{{\tt arXiv:1402.0364}}].

\bibitem{Botella:2014ska}
F.~Botella, G.~Branco, A.~Carmona, M.~Nebot, L.~Pedro, et~al., {\it {Physical
  Constraints on a Class of Two-Higgs Doublet Models with FCNC at tree level}},
   {\em JHEP} {\bf 1407} (2014) 078,
  [\href{http://arxiv.org/abs/1401.6147}{{\tt arXiv:1401.6147}}].

\bibitem{Joshipura:1990pi}
A.~S. Joshipura and S.~D. Rindani, {\it {Naturally suppressed flavor violations
  in two Higgs doublet models}},  {\em Phys.Lett.} {\bf B260} (1991) 149--153.

\bibitem{Gunion:2005ja}
J.~F. Gunion and H.~E. Haber, {\it {Conditions for CP-violation in the general
  two-Higgs-doublet model}},  {\em Phys.Rev.} {\bf D72} (2005) 095002,
  [\href{http://arxiv.org/abs/hep-ph/0506227}{{\tt hep-ph/0506227}}].

\bibitem{Ferreira:2009wh}
P.~Ferreira, H.~E. Haber, and J.~P. Silva, {\it {Generalized CP symmetries and
  special regions of parameter space in the two-Higgs-doublet model}},  {\em
  Phys.Rev.} {\bf D79} (2009) 116004,
  [\href{http://arxiv.org/abs/0902.1537}{{\tt arXiv:0902.1537}}].

\bibitem{Ferreira:2010hy}
P.~Ferreira, M.~Maniatis, O.~Nachtmann, and J.~P. Silva, {\it {CP properties of
  symmetry-constrained two-Higgs-doublet models}},  {\em JHEP} {\bf 1008}
  (2010) 125, [\href{http://arxiv.org/abs/1004.3207}{{\tt arXiv:1004.3207}}].

\bibitem{Bernon:2014vta}
J.~Bernon, B.~Dumont, and S.~Kraml, {\it {Status of Higgs couplings after Run-1
  of the LHC using Lilith 1.0}},  \href{http://arxiv.org/abs/1409.1588}{{\tt
  arXiv:1409.1588}}.

\bibitem{Coleppa:2013dya}
B.~Coleppa, F.~Kling, and S.~Su, {\it {Constraining Type II 2HDM in Light of
  LHC Higgs Searches}},  {\em JHEP} {\bf 1401} (2014) 161,
  [\href{http://arxiv.org/abs/1305.0002}{{\tt arXiv:1305.0002}}].

\bibitem{Eberhardt:2013uba}
O.~Eberhardt, U.~Nierste, and M.~Wiebusch, {\it {Status of the
  two-Higgs-doublet model of type II}},  {\em JHEP} {\bf 1307} (2013) 118,
  [\href{http://arxiv.org/abs/1305.1649}{{\tt arXiv:1305.1649}}].

\bibitem{Chen:2013rba}
C.-Y. Chen, S.~Dawson, and M.~Sher, {\it {Heavy Higgs Searches and Constraints
  on Two Higgs Doublet Models}},  {\em Phys.Rev.} {\bf D88} (2013) 015018,
  [\href{http://arxiv.org/abs/1305.1624}{{\tt arXiv:1305.1624}}].

\bibitem{Craig:2013hca}
N.~Craig, J.~Galloway, and S.~Thomas, {\it {Searching for Signs of the Second
  Higgs Doublet}},  \href{http://arxiv.org/abs/1305.2424}{{\tt
  arXiv:1305.2424}}.

\bibitem{Dumont:2014wha}
B.~Dumont, J.~F. Gunion, Y.~Jiang, and S.~Kraml, {\it {Constraints on and
  future prospects for Two-Higgs-Doublet Models in light of the LHC Higgs
  signal}},  {\em Phys.Rev.} {\bf D90} (2014) 035021,
  [\href{http://arxiv.org/abs/1405.3584}{{\tt arXiv:1405.3584}}].

\bibitem{Dev:2015bta}
P.~S.~B. Dev and A.~Pilaftsis, {\it {Natural Standard Model Alignment in the
  Two Higgs Doublet Model}},  \href{http://arxiv.org/abs/1503.09140}{{\tt
  arXiv:1503.09140}}.

\bibitem{Atlas:signal}
{\bf ATLAS} Collaboration, {\it {ATLAS-CONF-2014-009}},
  \href{http://arxiv.org/abs/https://atlas.web.cern.ch/Atlas/GROUPS/PHYSICS/CO%
NFNOTES/ATLAS-CONF-2014-009/}{{\tt
  https://atlas.web.cern.ch/Atlas/GROUPS/PHYSICS/CONFNOTES/ATLAS-CONF-2014-009%
/}}.

\bibitem{cms:signal}
{\bf CMS} Collaboration, {\it {CMS-PAS-HIG-14-009}},
  \href{http://arxiv.org/abs/http://cds.cern.ch/record/1728249?ln=en}{{\tt
  http://cds.cern.ch/record/1728249?ln=en}}.

\bibitem{Deshpande:1977rw}
N.~G. Deshpande and E.~Ma, {\it {Pattern of Symmetry Breaking with Two Higgs
  Doublets}},  {\em Phys.Rev.} {\bf D18} (1978) 2574.

\bibitem{Kastening:1992by}
B.~M. Kastening, {\it {Bounds from stability and symmetry breaking on
  parameters in the two Higgs doublet potential}},
  \href{http://arxiv.org/abs/hep-ph/9307224}{{\tt hep-ph/9307224}}.

\bibitem{Gunion:2002zf}
J.~F. Gunion and H.~E. Haber, {\it {The CP conserving two Higgs doublet model:
  The Approach to the decoupling limit}},  {\em Phys.Rev.} {\bf D67} (2003)
  075019, [\href{http://arxiv.org/abs/hep-ph/0207010}{{\tt hep-ph/0207010}}].

\bibitem{Klimenko:1984qx}
K.~Klimenko, {\it {On Necessary and Sufficient Conditions for Some Higgs
  Potentials to Be Bounded From Below}},  {\em Theor.Math.Phys.} {\bf 62}
  (1985) 58--65.

\bibitem{Maniatis:2006fs}
M.~Maniatis, A.~von Manteuffel, O.~Nachtmann, and F.~Nagel, {\it {Stability and
  symmetry breaking in the general two-Higgs-doublet model}},  {\em
  Eur.Phys.J.} {\bf C48} (2006) 805--823,
  [\href{http://arxiv.org/abs/hep-ph/0605184}{{\tt hep-ph/0605184}}].

\bibitem{Nie:1998yn}
S.~Nie and M.~Sher, {\it {Vacuum stability bounds in the two Higgs doublet
  model}},  {\em Phys.Lett.} {\bf B449} (1999) 89--92,
  [\href{http://arxiv.org/abs/hep-ph/9811234}{{\tt hep-ph/9811234}}].

\bibitem{Barroso:2013awa}
A.~Barroso, P.~Ferreira, I.~Ivanov, and R.~Santos, {\it {Metastability bounds
  on the two Higgs doublet model}},  {\em JHEP} {\bf 1306} (2013) 045,
  [\href{http://arxiv.org/abs/1303.5098}{{\tt arXiv:1303.5098}}].

\bibitem{Lee:1977eg}
B.~W. Lee, C.~Quigg, and H.~Thacker, {\it {Weak Interactions at Very
  High-Energies: The Role of the Higgs Boson Mass}},  {\em Phys.Rev.} {\bf D16}
  (1977) 1519.

\bibitem{Maalampi:1991fb}
J.~Maalampi, J.~Sirkka, and I.~Vilja, {\it {Tree level unitarity and triviality
  bounds for two Higgs models}},  {\em Phys.Lett.} {\bf B265} (1991) 371--376.

\bibitem{Kanemura:1993hm}
S.~Kanemura, T.~Kubota, and E.~Takasugi, {\it {Lee-Quigg-Thacker bounds for
  Higgs boson masses in a two doublet model}},  {\em Phys.Lett.} {\bf B313}
  (1993) 155--160, [\href{http://arxiv.org/abs/hep-ph/9303263}{{\tt
  hep-ph/9303263}}].

\bibitem{Akeroyd:2000wc}
A.~G. Akeroyd, A.~Arhrib, and E.-M. Naimi, {\it {Note on tree level unitarity
  in the general two Higgs doublet model}},  {\em Phys.Lett.} {\bf B490} (2000)
  119--124, [\href{http://arxiv.org/abs/hep-ph/0006035}{{\tt hep-ph/0006035}}].

\bibitem{Horejsi:2005da}
J.~Horejsi and M.~Kladiva, {\it {Tree-unitarity bounds for THDM Higgs masses
  revisited}},  {\em Eur.Phys.J.} {\bf C46} (2006) 81--91,
  [\href{http://arxiv.org/abs/hep-ph/0510154}{{\tt hep-ph/0510154}}].

\bibitem{Swiezewska:2012ej}
B.~Świeżewska, {\it {Yukawa independent constraints for two-Higgs-doublet
  models with a 125 GeV Higgs boson}},  {\em Phys.Rev.} {\bf D88} (2013), no.~5
  055027, [\href{http://arxiv.org/abs/1209.5725}{{\tt arXiv:1209.5725}}].

\bibitem{Pal:1994jk}
P.~B. Pal, {\it {What is the equivalence theorem really?}},
  \href{http://arxiv.org/abs/hep-ph/9405362}{{\tt hep-ph/9405362}}.

\bibitem{Horejsi:1995jj}
J.~Horejsi, {\it {Electroweak interactions and high-energy limit: An
  Introduction to equivalence theorem}},  {\em Czech.J.Phys.} {\bf 47} (1997)
  951--977, [\href{http://arxiv.org/abs/hep-ph/9603321}{{\tt hep-ph/9603321}}].

\bibitem{Das:2015qva}
D.~Das, {\it {New limits on $\mathbf{\tan\beta}$ for 2HDMs with $\mathbf{Z_2}$
  symmetry}},  \href{http://arxiv.org/abs/1501.02610}{{\tt arXiv:1501.02610}}.

\bibitem{Bhattacharyya:2013rya}
G.~Bhattacharyya, D.~Das, P.~B. Pal, and M.~Rebelo, {\it {Scalar sector
  properties of two-Higgs-doublet models with a global U(1) symmetry}},  {\em
  JHEP} {\bf 1310} (2013) 081, [\href{http://arxiv.org/abs/1308.4297}{{\tt
  arXiv:1308.4297}}].

\bibitem{Searches:2001ac}
{\bf LEP Higgs Working Group for Higgs boson searches, ALEPH, DELPHI, L3 and
  OPAL} Collaboration, {\it {Search for charged Higgs bosons: Preliminary
  combined results using LEP data collected at energies up to 209-GeV}},
  \href{http://arxiv.org/abs/hep-ex/0107031}{{\tt hep-ex/0107031}}.

\bibitem{He:2001tp}
H.-J. He, N.~Polonsky, and S.-f. Su, {\it {Extra families, Higgs spectrum and
  oblique corrections}},  {\em Phys.Rev.} {\bf D64} (2001) 053004,
  [\href{http://arxiv.org/abs/hep-ph/0102144}{{\tt hep-ph/0102144}}].

\bibitem{Grimus:2007if}
W.~Grimus, L.~Lavoura, O.~Ogreid, and P.~Osland, {\it {A Precision constraint
  on multi-Higgs-doublet models}},  {\em J.Phys.} {\bf G35} (2008) 075001,
  [\href{http://arxiv.org/abs/0711.4022}{{\tt arXiv:0711.4022}}].

\bibitem{Baak:2013ppa}
M.~Baak and R.~Kogler, {\it {The global electroweak Standard Model fit after
  the Higgs discovery}},  \href{http://arxiv.org/abs/1306.0571}{{\tt
  arXiv:1306.0571}}.

\bibitem{Ferreira:2009jb}
P.~Ferreira and D.~Jones, {\it {Bounds on scalar masses in two Higgs doublet
  models}},  {\em JHEP} {\bf 0908} (2009) 069,
  [\href{http://arxiv.org/abs/0903.2856}{{\tt arXiv:0903.2856}}].

\bibitem{Bezrukov:2012sa}
F.~Bezrukov, M.~Y. Kalmykov, B.~A. Kniehl, and M.~Shaposhnikov, {\it {Higgs
  Boson Mass and New Physics}},  {\em JHEP} {\bf 1210} (2012) 140,
  [\href{http://arxiv.org/abs/1205.2893}{{\tt arXiv:1205.2893}}].

\bibitem{Degrassi:2012ry}
G.~Degrassi, S.~Di~Vita, J.~Elias-Miro, J.~R. Espinosa, G.~F. Giudice, et~al.,
  {\it {Higgs mass and vacuum stability in the Standard Model at NNLO}},  {\em
  JHEP} {\bf 1208} (2012) 098, [\href{http://arxiv.org/abs/1205.6497}{{\tt
  arXiv:1205.6497}}].

\bibitem{Chakrabarty:2014aya}
N.~Chakrabarty, U.~K. Dey, and B.~Mukhopadhyaya, {\it {High-scale validity of a
  two-Higgs doublet scenario: a study including LHC data}},  {\em JHEP} {\bf
  1412} (2014) 166, [\href{http://arxiv.org/abs/1407.2145}{{\tt
  arXiv:1407.2145}}].

\bibitem{Das:2015mwa}
D.~Das and I.~Saha, {\it {Search for a 'stable alignment limit' in two
  Higgs-doublet models}},  \href{http://arxiv.org/abs/1503.02135}{{\tt
  arXiv:1503.02135}}.

\bibitem{Chowdhury:2015yja}
D.~Chowdhury and O.~Eberhardt, {\it {Global fits of the two-loop renormalized
  Two-Higgs-Doublet model with soft $Z_2$ breaking}},
  \href{http://arxiv.org/abs/1503.08216}{{\tt arXiv:1503.08216}}.

\bibitem{Bhattacharyya:2014oka}
G.~Bhattacharyya and D.~Das, {\it {Nondecoupling of charged scalars in Higgs
  decay to two photons and symmetries of the scalar potential}},  {\em
  Phys.Rev.} {\bf D91} (2015) 015005,
  [\href{http://arxiv.org/abs/1408.6133}{{\tt arXiv:1408.6133}}].

\end{thebibliography}\endgroup
